\documentclass[reprint,aps,amsmath,amssymb]{revtex4-1}
\usepackage[utf8]{inputenc} 
\usepackage[T1]{fontenc}    
\usepackage{hyperref}       
\usepackage{url}            
\usepackage{booktabs}       
\usepackage{amsfonts}       
\usepackage{nicefrac}       
\usepackage{microtype}      
\usepackage{float}
\usepackage{lipsum}
\usepackage{amsmath}
\usepackage{relsize}
\usepackage{graphicx}

\begin{document}
\title{Annihilation of single-species charged particles based on the Dyson gas dynamics}
\author{C.E.~Gonzalez-Ortiz} 
\email[ ]{gonza839@msu.edu}
\affiliation{Department of Physics and Astronomy, Michigan State University, East Lansing, MI 48824 USA}
\author{G.~Téllez}
\email[ ]{gtellez@uniandes.edu.co} 
\affiliation{Departamento de Física, Universidad de los Andes, Bogotá 111711, Colombia}

\begin{abstract}
We analyze the annihilation of equally-charged particles based on the Brownian motion model built by F. Dyson for $N$ particles with charge $q$ interacting via the log-Coulomb potential on the unitary circle at a reduced inverse temperature $\beta$, defined as $\beta=q^2/(k_B T)$. We derive an analytical approach in order to describe the large-$t$ asymptotic behaviour for the number density decay, which can be described as a power law, i.e., $n\sim t^{-\nu}$. For a sufficiently large $\beta$, the power law exponent $\nu$ behaves as $(\beta +1)^{-1}$, which was corroborated through several computational simulations. For small $\beta$, in the diffusive regime, we recover the exponent of 1/2 as predicted by single-species uncharged annihilation.   
\end{abstract}

\maketitle

\section{Introduction}
\label{sec:intro}
The annihilation of charged and uncharged particles has been a prolific area of interest for over 30 years now. The study of annihilation-diffusion kinetics of \textit{uncharged} particles has focused mainly on single-species reactions ($A+A\rightarrow \varnothing$) \cite{krapivsky_book,privman_2005,Ben_Naim_2016,racz1985,benav1986, ballistic2016} and two-species reactions ($A+B\rightarrow \varnothing$) \cite{krapivsky_book,privman_2005,amar2018,toussaint1983}. The study of annihilation-diffusion kinetics of \textit{charged} particles has focused primarily on the two-species opposite-charge reaction ($A_++A_-\rightarrow \varnothing$) interacting via a general long-range power-law attractive interaction, e.g. the Coulomb interaction \cite{burlatsky1996,ginzburg1997,kwon2006,jang1995,ispolatov1996}. For each of these reactions, a corresponding scaling theory has been formulated in order to explain the particle density decay and the critical dimensions of the systems \cite{toussaint1983,privman_2005,burlatsky1996}. All of this work has found relevance in several fields such as chemical reactions, fractal theory, topological defects, spin dynamics and superconductivity \cite{jang1995,Ben_Naim_2016}. Expanding on this entire framework, we propose a different approach in order to study single-species same-charge reactions, e.g., ($A_{+}+A_{+}\rightarrow \varnothing$), forced to interact in a close surface.            

Studying Coulomb systems allows scientists to gain important insight into complex systems, some of which at a microscopic scale interact via the Coulomb force. Some of these systems include plasmas, colloids and electrolytic solutions. In these systems, and specially in plasmas, it is of special interest studying how thermal fluctuations, electromagnetic and nuclear forces interact. If thermal fluctuations can overcome the electromagnetic repulsion between two nuclei, it allows the strong nuclear force to come into play and hold the two nuclei together. Hence, nuclear fusion occurs. This is a highly nonlinear and non-equilibrium problem. We propose a simple model system in order to explore these interactions. By studying the annihilation of single-species charged particles at low dimensionality we can gain insight as to how fusion occurs in such complex systems.   

We investigate the system of $N$ particles with charge $q$, at a reduced inverse temperature $\beta=q^2/(k_B T)$, interacting via the 2d-Coulomb log-potential and restricted to move around the unit circle. Additionally, when two particles are separated less than a critical fusion arc-length $\theta_f$, annihilation occurs for the pair of particles. We are specially interested in how the density $n$ decays over time. The configuration of $N$ charges restricted to move around the unit circle was extensively studied by F.~J.~Dyson \cite{dyson1962a,dysonI,dysonII,dysonIII}. In order to perform the simulations, we solve the Langevin equation for the system and compare the results with the Brownian model built by Dyson \cite{dyson1962a}. This approach allows us to explore the effect of $\beta$ on the density decay of the system. Other authors have used an Euler scheme \cite{ispolatov1996} and a lattice simulation \cite{kwon2006} in order to explore the annihilation of two-species charged particles while keeping $\beta$ constant. Our computational approach is similar to the one that Jang et al. \cite{jang1995} used to investigate two-species charged annihilation. In particular, they found that the density decays as $n(t)\sim t^{-\nu}$. Nevertheless, it is important to note that for two-species charged annihilation, the particles will be subjected to attractive and repulsive interactions, that ultimately accelerate annihilation, i.e., $\nu$ increases. For single species annihilation, the charges will only interact repulsively, which overall will slow down this process.

For our model and as we will show on our results, the power law exponent $\nu$, from $n(t)\sim t^{-\nu}$, will vary with respect to $\beta$, which differs greatly from the results for single-species uncharged annihilation and two-species charged annihilation. For sufficiently large $\beta$, the annihilation exponent $\nu$ will behave as $(\beta +1)^{-1}$. Nevertheless, as $\beta$ tends to zero, we recover the $\nu=1/2$ behavior predicted by single-species uncharged annihilation.   

In the next section of our paper, we will give an overview of the Dyson gas in thermal equilibrium, the solution to the Langevin equation for the Dyson gas out of equilibrium and some pertinent results from annihilation theory. In section \ref{sec:results}, we will show the results from our stochastic simulations. In section \ref{sec:discussion}, we propose a new approach using a modified kinetic rate equation in order to explain our results. And finally, in section \ref{sec:conclusion} we will make some concluding remarks. 

\section{Theory}
\label{sec:theory}

A Coulomb gas consists of $N$ charged particles interacting through
the $d$-Coulomb potential ($\varphi_d$), which reads
\begin{equation}
    \label{eq: dpotential}
     \varphi_d (\boldsymbol{r},\boldsymbol{r}') =\begin{cases}-q q'\,\, |\boldsymbol{r}-\boldsymbol{r}'|& d = 1\\-q q' \,\, \ln|\boldsymbol{r}-\boldsymbol{r}'| & d = 2 \\\mathlarger{-q q'|\boldsymbol{r}-\boldsymbol{r}'|^{-1}} & d=3 \\ \qquad ... \end{cases}, 
\end{equation}
depending on the number of dimensions $d$. The following work will focus on the $d=2$ case (two-dimensional), where a pair of particles located at $\boldsymbol{r}$ and $\boldsymbol{r}'$, with charge $q$ and $q'$ interact via a logarithmic potential with themselves and a neutralizing background. Nevertheless, the present work focuses specifically on the configuration where the $N$ charges are constrained to move on the unit circle. Therefore, the relevant variable will be the angular position ($\theta_i$) of each charged particle, hence, a 1-D problem.

For a pair of charged particles with labels $k$ and $j$, the corresponding position vectors, $\boldsymbol{r}_k$ and $\boldsymbol{r}_j$, can be defined by their angular coordinates $\theta_k$ and $\theta_j$, respectively. With this in mind, the quantity  $|\boldsymbol{r_j}-\boldsymbol{r_k}|$ just reads $|\exp{i\theta_j}-\exp{i\theta_k}|$. With the latter, this quantity can be introduced into the two-dimensional Coulomb potential, reading
\begin{equation}
    \label{eq: 2potential}
    \varphi_{2} (\theta_j,\theta_k)=-q^2 \ln\left({|\exp{i\theta_j}-\exp{i\theta_k}|}\right),
\end{equation}
where $q_j=q_k$ is set for single-species systems. Equation (\ref{eq: 2potential}) is the starting point of this work.

\subsection{The Dyson Gas}

The Dyson gas is a special type of Coulomb gas where $N$ classical, charged particles move on the unit circle. These charged particles exist within a neutralizing background, whose density $\rho_0$, is defined over the domain $\Omega$. For the sake of simplicity, $\Omega$ can be defined to be the unit circle, as well. Therefore, the density $\rho_0$ will just be a function of $\theta$, i.e. $\rho_0(\theta)$. With this in mind, the electroneutrality condition for the system just reads
\begin{equation}
    \label{eq:electroneutrality}
    \int _0 ^{2\pi} \rho_0 (\theta) d\theta=N.
\end{equation}

The simplest expression that satisfies this condition is
$\rho_0(\theta)=N/2\pi$. This background density corresponds to a
uniformly charged background. By taking the background into account,
the potential energy of the system $U_T(\theta_1,\dots,\theta_N)$ will
be the sum of charge-charge interaction, charge-background interaction
and background-background interaction, i.e.:
$U_T(\theta_1,\dots,\theta_N)=U_{CC}(\theta_1,\dots,\theta_N)+U_{BC}(\theta_1,\dots,\theta_N)+U_{BB}$.

The total charge-charge interaction is the sum over interactions of each pair of particles, reading: 
\begin{equation}
    \label{eq: cc}
    U_{CC}=-\sum _{j=1}^{N} \sum _{k=j+1}^N \ln \left({|\exp{i\theta_j}-\exp{i\theta_k}}|\right),
\end{equation}
where $q=1$, for simplicity. Secondly, the charge-background interaction is accounted for every particle's interaction with the entire background domain. This total charge-background potential energy reads:   
\begin{equation}
    \label{eq: cb}
    U_{BC}= \sum_{\ell=1}^N \int_0^{2\pi} \ln \left({|\exp{i\theta}-\exp{i\theta_{\ell}}}|\right) \rho_0(\theta) d\theta.
\end{equation}

Finally, the expression for the background-background potential reads:
\begin{equation}
    \label{eq: bb}
    U_{BB}=-\frac{1}{2}\int \rho_0(\theta) \rho_0(\theta') \ln|\exp{i\theta}-\exp{i\theta'}| d \theta' d\theta,
\end{equation}
where the factor of $1/2$ is taken into account in order to average
the interaction of the background with itself. Due to the
translation invariance on the unit circle, it can be shown that
$U_{BC}$ is constant \cite{forrester}. Therefore, when
working on the unit circle, the only relevant term for the potential
energy that is not constant is the charge-charge interaction $U_{CC}$.
Then, for a general configuration of charged particles, the
probability of finding $N$ charges at $\theta_1,\dots,\theta_N$ on the
unit circle at thermal equilibrium can be simplified to
\begin{equation}
    \label{eq: pdf}
    P(\theta_1,\dots,\theta_N)=\frac{1}{N! Z_N} \exp\left({-\beta U_{CC}(\theta_1,\dots,\theta_N)}\right),
\end{equation}
where $Z_N$ is the partition function that normalizes the probability.
Introducing equation (\ref{eq: cc}) into equation (\ref{eq: pdf}), yields:
\begin{equation}
    \label{eq: pdf1}
    P(\theta_1,\dots,\theta_N)=\frac{1}{N! Z_N} \prod_{\ell=1}^N \prod_{j=\ell+1}^N |e^{i\theta_j}-e^{i\theta_\ell}|^{\beta},
\end{equation}
after doing some manipulation on the sums and the exponential
terms. With equation (\ref{eq: pdf1}), the statistical physics of this
configuration can be studied directly. Additionally, it is noteworthy
that equation (\ref{eq: pdf1}), has the general structure of a
Vandermonde determinant, property that allows for appealing analytical
results \cite{mehta, forrester}. For instance, the partition function
\begin{equation}
  Z_N=\frac{1}{N!}\int_{[0,2\pi]^N}
  \prod_{\ell=1}^N \prod_{j=\ell+1}^N
  |e^{i\theta_j}-e^{i\theta_\ell}|^{\beta}
  d\theta_1\ldots d\theta_N
\end{equation}
can be related to a Selberg integral and be computed exactly for any
value of $\beta$ with the result~\cite{mehta, forrester}
\begin{equation}
  Z_N=\frac{(2\pi)^N}{N!} \frac{\Gamma(1+\beta
    N/2)}{(\Gamma(1+\beta/2))^N}
  \,
\end{equation}
where $\Gamma(x)=\int_0^{\infty} e^{-t} t^{x-1}\,dt$ is the gamma
function. Many of the thermal equilibrium properties of the Dyson gas
have been worked out explicitly and an account of those can be found
in the monographs by M.~L.~Mehta~\cite{mehta} and
P.~J.~Forrester~\cite{forrester}. In particular, it is worth noting
that the correlation functions have been computed exactly in terms of
generalized hypergeometric functions for even values of
$\beta$~\cite{forrester92, forrester93}.

In order to analyze the Dyson gas out of equilibrium, the Langevin equation for the system has to be stated. A general Langevin equation for the $k$-th particle confined to move in the unit circle, in the strong friction limit, has the following form
\begin{equation}
    \label{eq: lang1}
    f\dot{\theta_k}=-\frac{\partial W}{\partial \theta_k}+\eta (t),
\end{equation}
where $W$ corresponds to the interaction potential, independent of time,
acting on the particle and $f$ is a constant friction coefficient,
which fixes the rate of diffusion. In this case, $\eta(t)$ corresponds
to a random Gaussian white noise function, that complies with $\langle
\eta(t) \rangle=0$ and $\langle \eta (t) \eta (t') \rangle=2\beta^{-1}
\delta(t-t') $. The position of the particle would evolve according to
the following equation:
\begin{equation}
    \label{eq: langx}
    (\theta_k(t)-\theta_k(0)) f=\int_0^{t}\frac{\partial W}{\partial \theta_k} dt'+\int_0^{t}\eta(t')dt'.
\end{equation}
Therefore, for a small change $\delta \theta_k$, given a small time step $\delta t$, equation (\ref{eq: langx}) becomes:
\begin{equation}
    \label{eq: langx2}
    f\delta \theta_k=-\frac{\partial W}{\partial \theta_k}\delta t+\int_0^{\delta t}\eta(t')dt'.
\end{equation}
The general potential $W$ just takes the form of the charge-charge interaction stated in equation (\ref{eq: cc}), but it is important to keep in mind that we are working out of equilibrium. The potential $W$ reads
\begin{equation}
    \label{eq: W}
    W(\theta_1,\dots,\theta_N)=-\sum _{j,k}^{N}  \ln \left({|\exp{i\theta_j}-\exp{i\theta_k}}|\right).
\end{equation}
By taking the derivative of equation (\ref{eq: W}) with respect to $\theta_k$ and after doing some algebra, this yields
\begin{equation}
    \label{eq: circleforce}
    -\frac{\partial W}{\partial \theta_k}=\sum_{j\neq k} \frac{1}{2}\cot\left[ \frac{1}{2} (\theta_j-\theta_k) \right].
\end{equation}
The quantity described in equation (\ref{eq: circleforce}) just corresponds to the tangential force exerted on the $k$-th particle due to the interaction with the other $N-1$ charged particles. 

Introducing equation (\ref{eq: circleforce}) directly into equation (\ref{eq: langx2}) yields all the parameters needed in order to solve the Langevin equation for the Dyson gas confined to the unit circle. The next step is to take the expectation value of the equation in order to find the mean and the variance for the angular jumps $\delta \theta_k$ of a particle given a small $\delta t$. For this purpose, the special properties of the white noise function come into play. Therefore, the expected value for a small jump in the angular position for the $k$-th particle reads
\begin{equation}
    \label{eq: meancircle}
    \langle \delta \theta_k\rangle=\sum_{j\neq k} \frac{1}{2}\cot\left[ \frac{1}{2} (\theta_j-\theta_k) \right]\left(\frac{\delta t}{f} \right).
\end{equation}
The variance for this process yields
\begin{equation}
    \label{eq: varcircle}
    \langle (\delta \theta_k)^2 \rangle=\frac{2}{\beta}\left(\frac{\delta t}{f} \right).
\end{equation}
Dyson arrived to equations (\ref{eq: meancircle}) and (\ref{eq: varcircle}) by the means of perturbations to the circular ensemble of RMT \cite{dyson1962a}.

Our work focuses on using equations (\ref{eq: meancircle}) and (\ref{eq: varcircle}) to simulate the Brownian motion of the charges. With the addition that particles are subject to annihilation dynamics between them. Therefore, once a pair of particles are separated less than a critical fusion angle ($\theta_f$), an annihilation event occurs, and both particles are taken out of the simulation. In order for the system to remain electrically neutral, the background has to account for this sudden change. Some subtleties of the process behind this have to be stated. 

First of all, once the annihilation occurs we assume that the two charges fuse together to create a $+2q$-charged product, which will eventually merge with the background. Therefore, the reaction takes the form ($A_{+}+A_{+}\rightarrow B_{+2}^* \rightarrow \varnothing$), where the $B_{+2}^*$ product is an intermediate step to reach annihilation. Annihilation, in our context, means the coalescence of the $B_{+2}^*$ product with the neutralizing background. 

Therefore, the electroneutrality condition, as stated in equation (\ref{eq:electroneutrality}), will change. The new electroneutrality condition reads
\begin{equation}
    \label{eq:electroneutrality2}
    \int _0 ^{2\pi} \rho_0 (\theta;t) d\theta=N(t),
\end{equation}
where we have assumed the density $\rho_0$, changes with time, in order to account for the fusion/annihilation events. 

We will assume that the coalescence of the $B_{+2}^*$ product with the background happens in a time scale much faster than the diffusion dynamics. This means that we can ignore the interaction of the intermediate charged product with the other charges, and therefore, ignore its contribution to $U_{CC}$, as stated in equation (\ref{eq: cc}). The only change comes to the calculation of the background potential ($U_{BB}$), as defined in equation (\ref{eq: bb}), which will still remain virtually constant. $U_{BB}$ will just increase by a constant, every time there is a fusion/annihilation event. Since $U_{BB}$ and $U_{BC}$ are constants (they do not depend on the positions of the particles), they do not affect the forces on the particles. The dynamics of the system will not change if the background potential is held constant, after every annihilation event.

\begin{figure*}[ht]
\centering
\includegraphics[width=0.9\linewidth]{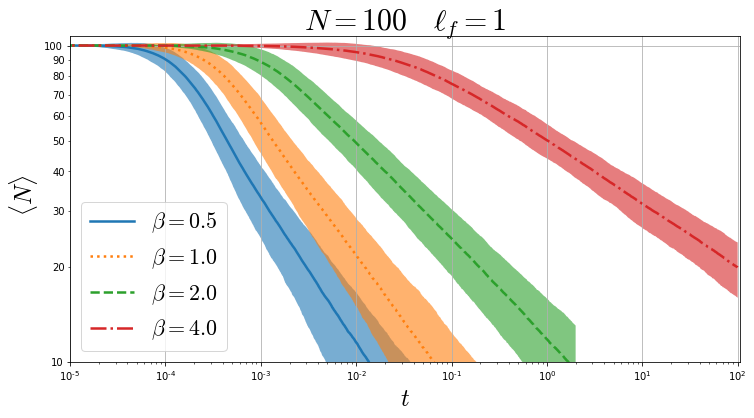}
\caption{Number of particles as a function of time for several temperatures ($\beta=$0.5, 1.0, 2.0 and 4.0) for $\ell_f=$1.0 and initial number of particles $N_0=$100. The simulations were done with $\delta t$ of $5\times10^{-8}$, $1\times 10^{-7}$, $1\times 10^{-7}$ and $4\times 10^{-6}$, respectively.}
\label{fig: N100LF1}
\end{figure*}

\subsection{Annihilation theory}

Single-species and two-species annihilation of uncharged particles is a well-known process, where down to a critical dimension ($d_{UC}$), the kinetic rate equation,
\begin{equation}
    \label{eq: kinetic_rate}
    \frac{dn}{dt}=-\mathcal{K} n^2,
\end{equation}
describes the density of the system. Equation (\ref{eq: kinetic_rate}) predicts that for large $t$, the density will decay as $n\sim t^{-1}$, in accordance with mean-field theory \cite{ginzburg1997,privman_2005,burlatsky1996}. For uncharged systems, when $d<d_{UC}$, diffusion and fluctuations start to become more relevant, thus slowing down the density decay. Specifically, for single-species reactions, $d_{UC}=2$. For these types of systems with dimensionality $d$ less than 2, the density decay slows down to
\begin{equation}
    \label{eq: single}
    n\sim t^{-d/2}.
\end{equation}
For our work, we are interested on the case $d=1$, corresponding to the unitary circle. Therefore, if there were no Coulomb interaction in our system, we would expect a decay exponent of $\nu=1/2$ in our results. 

Additionally, two-species annihilation can also be described by equation (\ref{eq: kinetic_rate}) in the mean-field region. Nevertheless, for these two-species systems, the upper critical dimension stands at $d_{UC}=4$. For these type of systems with dimensionality less than 4, the density decays as
\begin{equation}
    \label{eq: two}
    n\sim t^{-d/4}.
\end{equation}

Annihilation dynamics for systems of charged particles deviate significantly from the uncharged case, as described by equations (\ref{eq: kinetic_rate}), (\ref{eq: single}) and (\ref{eq: two}). The Coulomb interaction complicates the problem by adding additional correlations and long-range interactions to the mix. Charged particle annihilation has been mainly studied by describing the decay rate of the system of $N_0$ positive charges $+q$ and $N_0$ negative charges $-q$ interacting via the Coulomb potential, either in 1 dimension \cite{ispolatov1996} or 2 dimensions \cite{jang1995,ginz1995,yurke1993}, e.g., there was an extensive debate on the $\nu$ exponent for the 2D case.

Two-species charged annihilation, specifically the $A_++A_-\rightarrow \varnothing$ reaction, has been extensively studied \cite{jang1995,ginz1995,yurke1993,ispolatov1996}, and has found several applications under the framework of chemical reactions. The annihilation in this type of systems behaves as $n\sim t^{-v}$, in the scaling regime \cite{ginz1995}. The overall effect of the Coulomb interaction is to accelerate annihilation given the attraction between different species. In particular, Jang et. al. \cite{jang1995} studied the 2-d dimension case, for which equation (\ref{eq: two}) predicts an annihilation exponent of 1/2 for non-Coulombic annihilation. Jang et. al. found that in the strong diffusion limit, i.e., $\beta\ll 1$, the exponent is $0.55\pm0.05$, in accordance with 1/2. Nevertheless, in the deterministic regime, i.e., large $\beta$, $\nu$ is reported to be $0.90\pm0.05$.   

For our particular system, given that we worked with single-species charged particles, we expected no annihilation if diffusion was turned off ($\beta\rightarrow\infty$). Nevertheless, as diffusion became more relevant, as $\beta$ decreased, we expected annihilation to occur. Through the following work, we explored how diffusion and Coulomb forces interact affecting the annihilation dynamics of our system, and how different regimes can be identified.

\begin{figure*}[ht]
\centering
\includegraphics[width=0.9\linewidth]{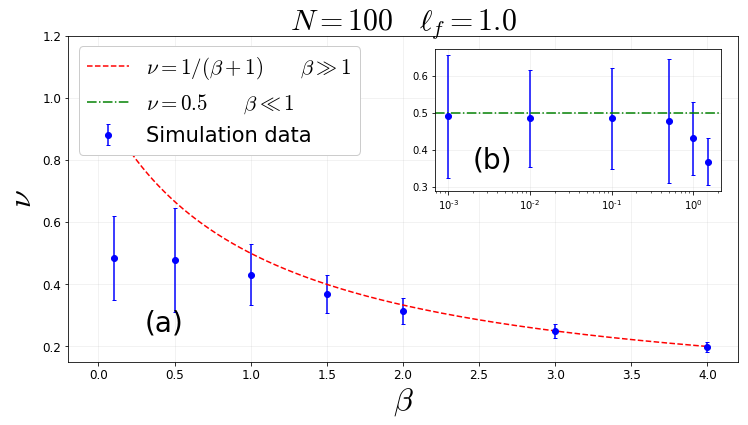}
\caption{Power law exponent of the density decay $\nu$ as a function of inverse temperature $\beta$. Red dashed line corresponds to theoretical behaviour as described in equation (\ref{eq: nsimt}). For large values of $\beta$ (a) the exponent matches the theoretical curve of $1/(\beta +1)$. For small $\beta$ (b), the exponent stabilizes near 1/2, green dash-dotted line.  }
\label{fig: NUvsBETA}
\end{figure*}

\section{Simulation details}
\label{sec:computational}

As stated in equations (\ref{eq: lang1}), (\ref{eq: meancircle}) and (\ref{eq: varcircle}), we used the Langevin equation based on the Dyson gas dynamics in order to analyze the annihilation dynamics of our system. Ultimately, the mean step for the Langevin equation up to a first order in $\delta t$ reads
\begin{equation}
    \label{eq: meanlang}
    \langle \theta_k(t+\delta t)-\theta_k(t)\rangle=\sum_{j\neq k} \frac{1}{2}\cot\left[ \frac{1}{2} (\theta_j-\theta_k) \right]\left(\frac{\delta t}{f} \right),
\end{equation}
and the variance for this stochastic process is
\begin{equation}
    \label{eq: varlang}
    \langle \Delta ( \theta_k(t+\delta t)-\theta_k(t))^2 \rangle=\frac{2}{\beta}\left(\frac{\delta t}{f} \right).
\end{equation}

We ran simulations for an initial number of charges ($N_0=100$) evolving the system stochastically every time step according to equations (\ref{eq: meanlang}) and (\ref{eq: varlang}). Every time step the simulation checked for particles that were separated less than a critical fusion angle ($\theta_f$). The definition for the critical fusion angle ($\theta_f$) reads:
\begin{equation}
    \label{eq: thetacritical}
    \theta_f=\left(\frac{\ell_f}{10}\right)\frac{2\pi}{N_0}=\left(\frac{\ell_f}{10}\right)S_{N_0},
\end{equation}
where $S_{N_0}$ corresponds to the initial mean separation of the system, i.e.: $S_{N_0}=2\pi/N_0$. If the separation of two particles is less than our parameter $\theta_f$, the two charges are taken out of the system. The parameter $\ell_f$ is just defined for convenience, in order to illustrate the fraction of the initial separation used as critical fusion angle. This parameter $\ell_f$ is changed for the simulations. The number of charges left in the system is monitored every time iteration. Each simulation consisted of over $10^6$ time steps, where the latter scheme was applied. 

The time step $\delta t$ was chosen to be such that the diffusive displacement was much less than the critical mean separation of the system as encoded in $\theta_f$. Thus, we constrained equation (\ref{eq: varlang}) in order to account for any undesired jump between particles without annihilation. Mathematically, this constrain can be expressed as $\sigma\ll\theta_f$, where $\sigma$ is the standard deviation of the Langevin process, as defined in equations (\ref{eq: meanlang}) and (\ref{eq: varlang}). Solving this constrain for $\delta t$, we get an additional constrain on $\delta t$ for the simulations. This constrain reads 
\begin{equation}
\label{eq: dtcritical}
    \delta t \ll \frac{2 \pi ^2 \beta f}{N_0^2}\left( \frac{\ell_f}{10}\right)^2,
\end{equation}
where we chose $f=1$ in order to absorb the friction parameter into the virtual time units.

With the computer simulation details sorted out, we performed at least 1000 simulations per intersection of parameters $\ell_f$ and $\beta$. Specifically, we ran simulations for $\ell_f=0.5$ to $\ell_f=8.0$, where we explored $\beta$ from 0.01 to 4. For higher $\beta$, the simulations were computationally unfeasible given that the computational time was too large in order to report our data with satisfactory statistics.    

\begin{figure*}[ht]
\centering
\includegraphics[width=0.9\linewidth]{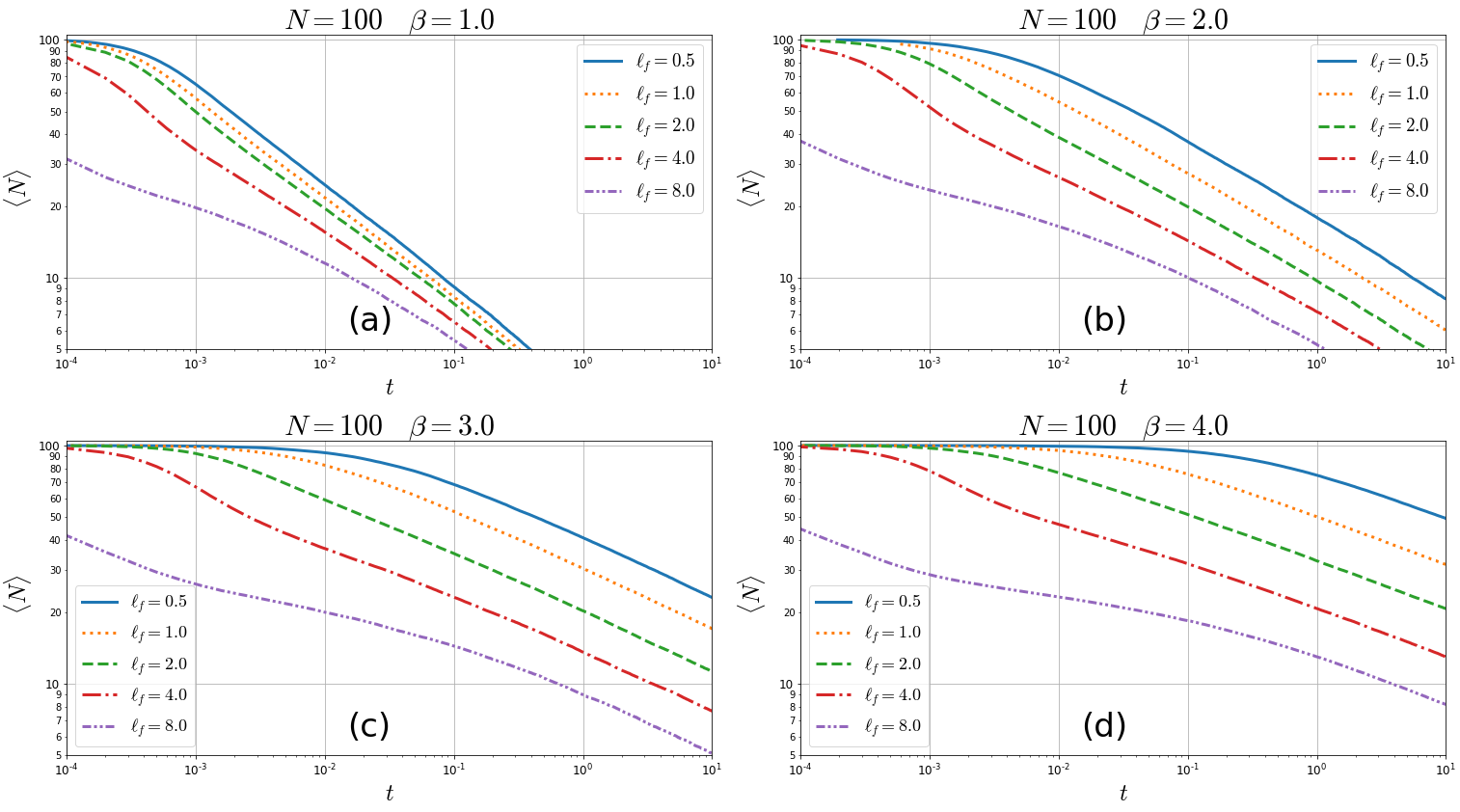}
\caption{Number of particles as a function of time for several fusion parameters ($\ell_f=$0.5, 1.0, 2.0, 4.0 and 8.0) at (a) $\beta=$1.0, (b) $\beta=$2.0, (c) $\beta=$3.0, and (d) $\beta=$4.0, for an initial number of particles of $N_0=$100. Simulations were done with a $\delta t$ of $1\times 10^{-5}$.}
\label{fig: N100B4}
\end{figure*}

\section{Results}
\label{sec:results}

All the simulations began with a fixed number particles, $N_0=100$. Every time step the number of particles was monitored in order to analyze the annihilation dynamics of the system. Figure \ref{fig: N100LF1} shows the number of particles for several $\beta$ as a function of time. It is clear how the particles start to annihilate themselves as the system evolves. Furthermore, for sufficiently long times, the system would annihilate completely, with no particles remaining, even though the Coulomb force was still acting. 

Figure \ref{fig: N100LF1} also shows the stochastic nature of our approach. The shaded regions around the curves represent the fluctuations around the mean number of particles for that time step, i.e., the lower and upper bound correspond to $\langle N \rangle-2\sigma_N$ and  $\langle N \rangle+2\sigma_N$, where $\sigma_N$ is defined as the standard deviation for the number of particles in the system at a time $t$. This means that we have approximately 95.5\% of the simulations included inside the shaded region. These fluctuations come from the fact that at least 1000 stochastic simulations were done for each parameter configuration. Figure \ref{fig: N100LF1} also shows how these fluctuations decrease as $\beta$ increases. This behaviour can be seen more clear on figure \ref{fig: NUvsBETA}. As $\beta$ increases, diffusion starts playing a weaker role in the annihilation dynamics, as compared to the Coulomb force. 

The annihilation rate of the system decreases as $\beta$ increases. This can be seen on figure \ref{fig: N100LF1}, where for $\beta=4.0$ there were still particles in the system after $t=10^1$, whereas for $\beta=0.5$, all particles were annihilated near $t=10^{-2}$. Furthermore, the annihilation dynamics follow a power law, i.e. $n\sim t^{-\nu}$, for large $t$. From a plot such as the ones shown in figure \ref{fig: N100LF1}, we can calculate the $\nu$ exponent by obtaining the slope of the line after an arbitrarily large $t_0$. The uncertainty of $\nu$ was calculated by applying error propagation on the regression analysis. This way, we calculated the annihilation exponent and its uncertainty for several $\beta$ and $\ell_f$ given $N_0=100$. This information is synthesized in figures \ref{fig: NUvsBETA} and \ref{fig: N100B4}.

Figure \ref{fig: NUvsBETA} shows the behaviour of the $\nu$ exponent, or annihilation exponent, with respect to the inverse temperature $\beta$ of the simulated system. Two asymptotic regimes can be explored. The first one corresponds to the limit case where $\beta \rightarrow 0$, i.e., where the temperature is so large that diffusion dynamics becomes dominant. In this case, as can be seen in figure \ref{fig: NUvsBETA}, specifically in part (b), the $\nu$ exponent tends to 1/2. This value corresponds to the annihilation exponent if there were no log-Coulomb interaction, only diffusion driving the annihilation process. Therefore, we expected this $\nu=1/2$ behavior for small $\beta$, given that diffusion would eventually outweigh the Coulomb forces for sufficiently small $\beta$. Overall, as $\beta \rightarrow 0$, diffusion drives the annihilation dynamics of our system.      

The second regime corresponds to the limit case where $\beta \gg 1$, i.e., where the log-Coulombic forces outweigh the diffusion kinematics. In this regime, for sufficiently large $\beta$, the $\nu$ exponent behaves as $1/(\beta+1)$. The origin of this behaviour will become clear after the theoretical discussion proposed in the next section. Nevertheless, it is important to note from figure \ref{fig: NUvsBETA} that the annihilation rate of our system will continue to decrease as $\beta$ increases, or, equivalently, as the temperature decreases. This means that for sufficiently large $\beta$, the annihilation exponent will eventually tend to 0. Virtually, for these small temperatures annihilation would not occur, given that diffusion is no longer strong enough to drive this process through.   

Another important thing we can note from figure \ref{fig: NUvsBETA} are the uncertainties reported for the $\nu$ exponent, or annihilation exponent. As mentioned previously, the number of particles' fluctuations was significantly larger for smaller $\beta$. Nevertheless, as these fluctuations became higher, the uncertainty on our annihilation exponent also increased. On the other hand, as $\beta$ increased, the uncertainty on the exponent decreased. This goes hand-to-hand with the argument that at high temperatures, annihilation is driven by thermal diffusion, while at low temperatures, annihilation is controlled by the log-Coulomb interaction of our system.

Figures \ref{fig: N100LF1} and \ref{fig: NUvsBETA} show how the annihilation dynamics depends on the inverse temperature $\beta$ of our system. Nevertheless, there is another parameter that regulates this annihilation. We also expected the fusion parameter $\theta_f$, as defined in equation (\ref{eq: thetacritical}), to play a role in the annihilation dynamics. Figure \ref{fig: N100B4} and its subsequent subfigures show the average number of particles as a function of time for several $\ell_f$ given a $\beta$. The parameter $\theta_f$ represents the angle of separation at which a pair of charged particle annihilate. Therefore, if $\theta_f$ is increased ---or equivalently, if $\ell_f$ is increased--- we expected the annihilation of charges to accelerate. Figures \ref{fig: N100B4}(a), \ref{fig: N100B4}(b), \ref{fig: N100B4}(c), and \ref{fig: N100B4}(d) show how for larger values of $\ell_f$, the charged particles consume themselves more rapidly.

For large values of $\ell_f$ ($\ell_f=4, 8$), the system exhibits two different time regimes. First, an initial transient regime with a very fast annihilation, which is simply due to the fact that the average spacing between particles is comparable to the fusion arc-length $\theta_f$. Therefore, there will be many frequent annihilation events due to the fact that there is not enough place to move for the particles before they encounter another one to annihilate with. After this transient regime, the density will decay enough to give more room for the particles to move. The system then reaches the scaling regime, where $n\sim t^{-\nu}$, in which $l_f$ does not affect the decay exponent $\nu$, only the proportionality constant in this power-law decay. 
This is illustrated in figure \ref{fig: N100B4} which shows how, for large times, the charged-particle systems, given different $\ell_f$, annihilate with the same $\nu$ exponent. Specifically, figure \ref{fig: N100B4}(b) shows how the curves become parallel to each other after $t=10^{-1}$, which means that they have the same slope given by the same $\nu$ exponent. That is to say that the annihilation exponent of the system does not depend on $\theta_f$ for the large-$t$ asymptotic behaviour. Therefore, the $\nu$ annihilation exponent will behave as $1/(\beta+1)$, independent of the value of $\theta_f$ for large $\beta$. 

Based on this evidence from the simulations, we can formulate an empiric formula for the density of the system as $t\rightarrow\infty$. This equation goes as:
\begin{equation}
    n(t)=\Phi_F(\theta_f,\beta)t^{-\nu},
    \label{eq: empiric}
\end{equation}
where $\Phi_F$ is just a constant dependent on $\beta$ and $\theta_f$. Equation (\ref{eq: empiric}) arises naturally when we analyze the simulated data. Nevertheless, on the next section, we propose an analytic approach based on Wigner's surmise \cite{mehta} on level separation, in order to get a similar expression.

Equation (\ref{eq: empiric}) encapsulates the time dependence of the annihilation, nonetheless, additional information can be gathered about $\Phi_F$ from the simulations. By taking the logarithm on both sides of equation (\ref{eq: empiric}), we get
\begin{equation}
    \ln n=-\nu \ln t+\ln \Phi_F.
    \label{eq: logemp}
\end{equation}
By taking the slope of the linear regression between $\ln n$ and $\ln t$, we were able to get the values of the annihilation exponent, just as explained before, and just as figure \ref{fig: NUvsBETA} shows. By doing this, we also gathered information on the dependence of $\Phi_F$ on $\theta_f$ and $\beta$. For simplicity, the new constant $C_F$ is defined as $C_F=\ln \Phi_F$.

\begin{figure}[h]
\centering
\includegraphics[width=\linewidth]{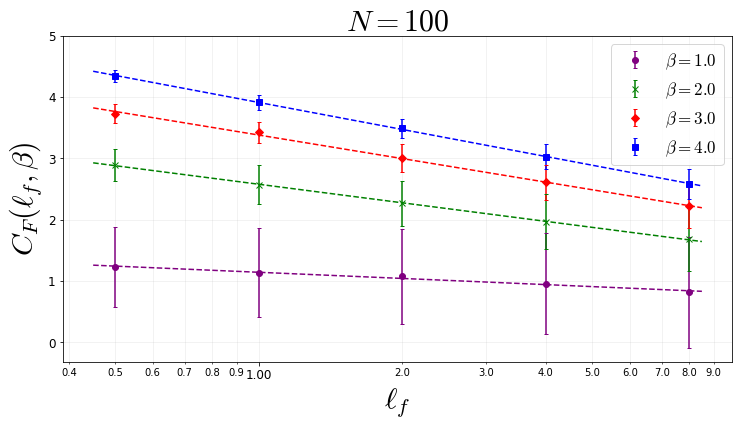}
\caption{The value of the regression constant $C_F$ plotted against $\ell_f$ for several values of $\beta$.}
\label{fig: Cfvslf}
\end{figure}

Figure \ref{fig: Cfvslf} shows the values of the regression constant $C_F$ plotted against the fusion parameter $\ell_f$ for the four temperatures in figure \ref{fig: N100B4}. One thing that becomes clear, is the increase of the fluctuations of the system as $\beta$ decreases. This can be seen by analyzing the size of the error bars, which represent the fluctuations of the system. Additionally, figure \ref{fig: Cfvslf} shows how $C_F$ depends on the parameter $\ell_f$ elevated to some power that depends on $\beta$. This is equivalent to saying that the constant $\Phi_F$ goes as $\Phi_F\sim (\theta_f)^\gamma$, where the $\gamma$ exponent is dependent on $\beta$. This means that the density decay behaves as 
\begin{equation}
    \label{eq: thetadecay}
    n(t)\sim (\theta_f)^\gamma t^{-\nu},
\end{equation}
for all $\beta$. Nevertheless, for small $\beta$, we can see from figure \ref{fig: Cfvslf}, how the constant $C_F$ becomes independent of the fusion parameter $\theta_f$ of the system. For $\beta=1$ in figure \ref{fig: Cfvslf}, the plot of $C_F$ starts to become horizontal, which means a very small variation of $C_F$ when $\ell_f$ is changed. Although smaller values of $\beta$ are not plotted, we discovered this phenomena to be true for $\beta\ll1$. In the diffusive regime, the arbitrary fusion length $\theta_f$ plays no role in the large-$t$ annihilation dynamics, as long as the condition in equation (\ref{eq: dtcritical}) is met.

\begin{figure}[h]
\centering
\includegraphics[width=\linewidth]{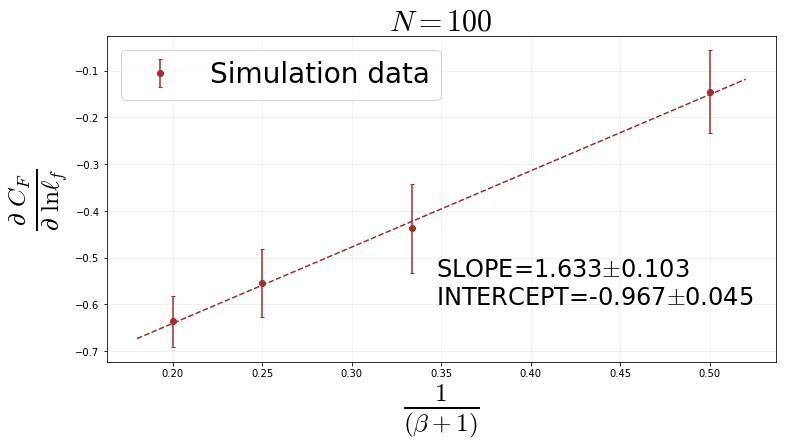}
\caption{The slopes of the curves from figure \ref{fig: Cfvslf} plotted as a function of $1/(\beta +1)$ with the corresponding linear regression for the data.}
\label{fig: Cfvsbetainv}
\end{figure}

On the other hand, we can explore the behavior of $\Phi_F$ for sufficiently large values of $\beta$. Figure \ref{fig: Cfvsbetainv} shows the slope values with its correspondent uncertainty from the plots in figure \ref{fig: Cfvslf}. After some analysis, figure \ref{fig: Cfvsbetainv} tells us that the constant $C_F$ will go as $C_F\sim (A/(\beta+1)+B)\ln \ell_f$ for sufficiently large $\beta$. By taking into account the definition of $\theta_f$ from equation (\ref{eq: thetacritical}) and the definition of $C_F$, we can rewrite this behavior as    
\begin{equation}
    \Phi_F\sim(\theta_f)^{\frac{A}{\beta+1}+B}.
    \label{eq: phif}
\end{equation}
In particular, if we take the limit $\beta\gg 1$, equation (\ref{eq: phif}) tells us that $\Phi_F\sim(\theta_f)^{B}$. From the linear regression in figure \ref{fig: Cfvsbetainv}, we have that $B=-0.97\pm0.05$ and $A=1.63\pm0.10$.  

\begin{figure}[h]
\centering
\includegraphics[width=\linewidth]{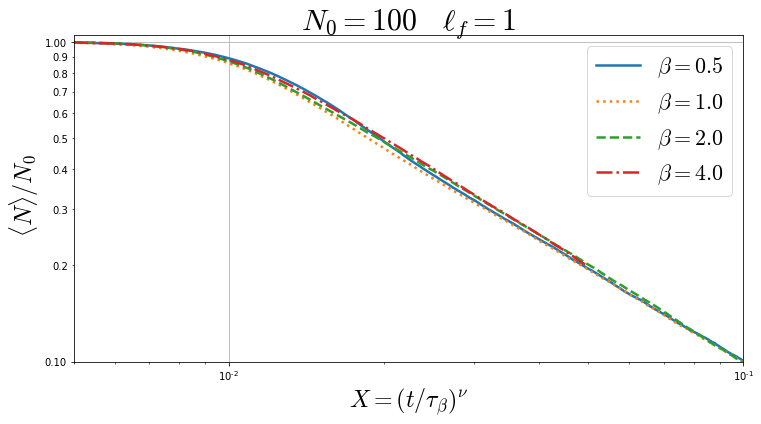}
\caption{Master curve for the graphs shown in figure \ref{fig: N100LF1}. Plots were collapsed using a change of variables with the correct $\tau_\beta$, for each curve.}
\label{fig: master}
\end{figure}

In summary, the data has shown that the annihilation of single-species charged particles will behave as 
\begin{equation}
    \label{eq: summary1}
    n(t) \sim (\theta_f)^{\frac{A}{\beta+1}+B} t^{-\nu},
\end{equation}
for large-t asymptotics and sufficiently large $\beta$. We can go one step further by summarizing the scaling behaviour of the system, and collapsing the curves into one master curve, as shown in figure \ref{fig: master}. We can make a change of variables such that $t \rightarrow X=(t/\tau_\beta)^\nu$, in order to collapse all the curves. It is clear that $\nu$ and $\tau_\beta$, are calculated from the simulations, and will change for each $\beta$ and $\ell_f$. The resulting master curve has a slope of $-1$, as shown in figure \ref{fig: master}. This means that the fraction of particles in the system go as: 
\begin{equation}
    \label{eq: master}
    Y\equiv\frac{N(t)}{N_0}=X^{-1}.
\end{equation}

\section{Discussion}
\label{sec:discussion}

The last section showed how the results from the simulations follow a quantifiable trend. On this section, these behaviors are derived based on physical arguments starting from the kinetic rate equation. In particular, we are interested in the annihilation dynamics for $\beta\gg 1$ regime, where the Coulomb interaction is the dominant interaction.

The kinetic rate equation, equation (\ref{eq: kinetic_rate}), is based on mean-field theory arguments. This equation doesn't take into account the correlations between particles, and, thus, breaks down for sufficiently small dimensions. In our case, given the log-Coulomb interaction, we need to include a term to account for the repulsion between particles. This term is equal to the probability of finding two particles spaced less than our critical fusion angle $\theta_f$. Based on these arguments, we can write an annihilation rate equation for our system, which reads:     
\begin{equation}
\label{eq: dndt}
    \frac{dn}{dt}=-\alpha_0\ n \left( \int^{n\theta_f}_0 p(s,t) ds \right),
\end{equation}
where $p(s,t)ds$ is the probability distribution function for the spacing $s$ between particles, and will, in general, depend on time. This spacing $s$ is defined as $s=n\theta$, where after a long time without annihilation, $\langle s\rangle=1$. This means that after a long time, the charges, on average, will be distributed equispatially on the circle. Therefore, if annihilation is slow enough, spacing distribution will be independent of time, i.e., $p(s,t)ds=p(s)ds$. For constant $\theta_f$, annihilation will occur slowly when $\beta\gg1$. Therefore, the approximation $p(s,t)=p(s)$ will only be valid for large $\beta$, where the system has enough time to accommodate to the equilibrium configuration, without the annihilation of particles. 

The explicit functional forms of $p(s)ds$ for several $\beta$ are calculated explicitly in \cite{mehta}. Nevertheless, a good approximation for these functions comes from Wigner's surmise, which reads:  
\begin{equation}
\label{eq: ps}
    p(s)=K s^{\beta}\exp[{-\zeta s^2}],
\end{equation}
where for sufficiently small spacings ($s$), this equation can be approximated to:
\begin{equation}
\label{eq: ps0}
    p(s)\approx K s^{\beta}.
\end{equation}
If we introduce equation (\ref{eq: ps0}) into equation (\ref{eq: dndt}) and solve the integral, we get a differential equation for the annihilation dynamics of our system, which reads:
\begin{equation}
\label{eq: dndt3}
    \frac{dn}{dt}=-\left( \frac{\alpha \ \theta_f^{\beta +1}}{\beta +1}\right) n^{\beta+2},
\end{equation}
where $\alpha=K\alpha_0$. We can go ahead and solve equation (\ref{eq: dndt3}), in order to get an expression for the temporal dependence of the density in our system. The entire solution reads: 
\begin{equation}
\label{eq: nt1}
    n\sim \theta_f^{-1}(\alpha t)^{-\frac{1}{\beta +1}},
\end{equation}
where the most important result is the power law on time, where the scaling exponent is dependent on $\beta$. Therefore, according to these arguments, the asymptotic behavior of the density of our system for large $t$ and large $\beta$ goes as:
\begin{equation}
\label{eq: nsimt}
    n \sim t^{-\frac{1}{\beta +1}}.
\end{equation}
This behavior can be compared to equations (\ref{eq: single}) and (\ref{eq: two}) for single species uncharged annihilation, where the exponent does not depend on $\beta$. Furthermore, this power law behavior is the same as the one shown in figure \ref{fig: NUvsBETA}, which accounts for the data trend in the simulations.

Equation (\ref{eq: nt1}) also tells us the density dependence on the critical fusion angle $\theta_f$. The density goes as $n(t)\sim \theta_f^{-1}$, for sufficiently large $t$ and $\beta$. If we compare this expression with equations (\ref{eq: phif}) and (\ref{eq: summary1}) in the limit case where $\beta \gg 1$, we see the agreement with the simulated data. In particular, the regression intercept from figure \ref{fig: Cfvsbetainv}, B$=-0.97\pm0.05$, is in agreement with the $-1$ exponent predicted by our analysis. 

\section{Conclusion}
\label{sec:conclusion}

A computational model based on the Dyson gas dynamics was developed in order to explore the annihilation of equally-charged particles forced to interact in a circumference through a logarithmic potential. The model simulated through Brownian dynamics how the particles annihilated given a critical fusion parameter $\theta_f$ and an inverse temperature $\beta$. We were able to explore the annihilation dynamics of our system, and in particular we were able to characterize the density for large times through a power law, i.e., $n\sim t^{-\nu}$. 

Through the simulations, we were able to distinguish three different long-time annihilation scenarios in our system. The first scenario corresponds to diffusive annihilation, where $\beta\ll1$. For this case, we observed that the annihilation exponent $\nu$ tended to $1/2$, just as previous literature indicated for 1-D annihilation  \cite{privman_2005,toussaint1983}. The second scenario corresponds to Coulomb annihilation, where $\beta\gg 1$. In this case, the $\nu$ exponent behaves as $1/(\beta+1)$. Additionally, we proposed an analytical approach that explains the physical origin of this behavior based on a modified kinetic rate equation that accounts for the Coulomb correlations between particles. The third scenario corresponds to intermediate values of $\beta$ where both diffusion and Coulomb forces mediate the annihilation dynamics of our system. These annihilation scenarios can be seen clearly on figure \ref{fig: NUvsBETA}.

In a general framework, our work built an approach to investigate single-species charged particle annihilation under a logarithmic potential. Nevertheless, this work focused on the 1-D case. We know from uncharged annihilation processes that this phenomena depends on the dimensionality of the inspected system. Therefore, it is of special interest to characterize the annihilation of charged particles at higher dimensions. For single-species uncharged annihilation we know that the critical dimension of the system is $d_{UC}=2$ \cite{burlatsky1996}, i.e., the dimension at which mean-field theory breaks down. It is still unclear as to how dimensionality plays a role on charged particle annihilation for large $\beta$.         

\section*{Acknowledgments} 
G.T. acknowledges support from Fondo de
Investigaciones, Facultad de Ciencias, Universidad de los Andes,
Research Program 2018-2019 ``Modelos de baja dimensionalidad de
sistemas cargados'' and ECOS-Nord project C18P01. We would also like to thank HPC Uniandes for providing high performance computing time.

\bibliographystyle{unsrt}
\bibliography{references}

\end{document}